\documentclass[preprint,showpacs,preprintnumbers,amsmath,amssymb]{revtex4}

\usepackage{graphicx}% Include figure files
\usepackage{bm}% bold math
\begin{document}

\title{Can coarse-graining introduce long-range correlations in a symbolic sequence?}
\author{S. L. Narasimhan, Joseph A. Nathan$^*$ and K. P. N. Murthy$^{**}$}
\affiliation{Solid State Physics Division, Bhabha Atomic Research Center, Mumbai-400085, India \\ $^*$Reactor Physics Design Division, Bhabha Atomic Research Center, Mumbai-400085, India \\ $^{**}$Materials Science Division, Indira Gandhi Center for Atomic Research, Kalpakkam - 603102, Tamilnadu, India
}
\vspace{3cm}

\begin{abstract}
We present an exactly solvable mean-field-like theory of correlated ternary
sequences which are actually systems with two independent parameters. Depending on the values of these parameters, the variance on the average number of any given symbol shows a linear or a superlinear dependence on the length of the sequence. We have shown that the available phase space of the system is made up a diffusive region surrounded by a superdiffusive region. Motivated by the fact that the diffusive portion
of the phase space is larger than that for the binary, we have studied the mapping
between these two. We have identified the region of the ternary phase space, particularly the
diffusive part, that gets mapped into the superdiffusive regime of the binary. This exact 
mapping implies that long-range correlation found in a lower dimensional representative 
sequence may not, in general, correspond to the correlation properties of the original system.
\end{abstract}
\maketitle

The dynamical behavior of complex systems consisting of a large number of hierarchically
organized subsystems is known to carry the signature of long-range spatio-temporal 
correlations (LRC) that might be present in them. A variety of physical [1], biological [2], linguistic [3] and even financial systems [2] provide examples of LRC systems. A standard method for studying correlations in such a system is to first divide its state space into a finite number of distinctly labelled regions, map the sequence of states assumed by the system into a sequence of these symbols and then study the statistical properties of this representative sequence. 

Since such a {\it coarse-graining} procedure is not expected to lead to a loss of long-range correlations in the system, we may choose to divide the state-space into two distinct regions and study the statistical properties of the resulting binary sequence. In fact, we generate a correlated binary sequence, parametrizable by a known procedure, and try to find the parameter values that correspond to the representative sequence.

Recently, it has been shown [4] that the strength of long-range correlations in a binary sequence of length, $N$, can be characterized by a parameter $\mu \in [-1,1]$ so that
non-trivial correlations between the symbols correspond to the parameter region, 
$\mid\mu \mid > 1/2$. More specifically,  the variance, $\sigma ^2 (N)$, of the the number of a symbol has been shown to have a form, $\sigma ^2 (N)\propto N^\alpha$, where $\alpha = 1$ for $\mid\mu \mid \leq 1/2$ ({\it diffusive }) and $\alpha > 1$ for $\mid\mu \mid > 1/2$ ({\it superdiffusive}) [5]. This exact mean-field-like theory of correlated binary sequences seems to provide a paradigm for studying the correlational properties of generic symbolic sequences such as even natural language texts. 

The implicit assumption in this approach is that an LRC system can be represented by a correlated binary sequence. There may be no {\it a priori} reason why this assumption should hold good. For example, it has been argued [6] that we need a minimum of ten letters (not less than five letters, in any case) to be able to design a foldable model of amino acid sequences. 
So, the minimum number of symbols required for a sequential representation of the system 
(or equivalently, the extent to which the state-space of a system can be coarse-grained) may
depend on the specific behaviour of the system under study. Even if a binary sequential representation is acceptable, the problem of how to arrive at this representation remains if the number of symbols originally associated with the system is odd. 

Thus, it is necessary to examine whether a coarse-graining procedure could introduce
long-range correlation, besides the one that might be present in the sequence already [7].
To this end, we present an exact mean-field-like theory of correlated ternary sequences
and identify the diffusive subregion of its phase space that gets spuriously mapped into the superdiffusive region of the phase space associated with the binary sequence. This exact 
mapping implies that long-range correlation found in a lower dimensional representative 
sequence may not, in general, correspond to the correlation properties of the original system.

\section{Ternary Sequences} 
Let $T_{N,i}\, (\, \equiv t_{i-N}t_{i-N+1},...,t_{i-1})$ 
denote a subsequence of $N$ ternary symbols, $t \in \{ 0,1,2\} $. Then the conditional probability, $p(t_i \mid T_{N,i})$, that the $i$th symbol $t_i$ in the sequence will be 
$0,1$ or $2$ may not only depend on the number of individual symbols but, in general, may also depend on their specific order in $T_{N,i}$. Ignoring the configuration-dependence of $p(t_i \mid T_{N,i})$ leads to a solvable mean-field-like theory of these ternary sequences. We define the conditional probability, $p(t_i = 0 \mid T_{N,i})$ as follows:
\begin{equation}
p(t_i = 0 \mid T_{N,i}) \equiv p_0 (T_{N,i}) = \frac{1}{N} (n_0 g_0 + n_1 g_1 + n_2 g_2)
\end{equation}
where $n_0$, $n_1$ and $n_2$ denote the number of $0$'s, $1$'s and $2$'s respectively in the sequence such that $n_0 + n_1 + n_2 = N$, and $g_0$, $g_1$ and $g_2$ denote the {\it a priori} probabilities of choosing the respective symbols of which only two are
{\em independent} because $g_0 + g_1 + g_2 = 1$. We can parametrize the deviations of $g_t$ from their 'unbiassed' values $1/3$ by writing $g_t = (1 + \mu _t)/3$ where
$\mu _{t = 0,1,2} \in [-1,2]$ and $\mu _0 + \mu _1 + \mu _2 = 0$. It is clear that the $\mu$'s
are a measure of the 'memory' built into the system which in turn leads to 
correlations between the symbols.

Since one of the three symbols is definitely going to be found at any given place in the subsequence, {\it i.e.,}$\sum _t p_t (T_{N,i}) = 1$, we need only to define $p_1 (T_{N,i})$ 
or $p_2 (T_{N,i})$. Now, the probability that the $i^{th}$ symbol is {\it not} $0$ is given by
\begin{equation}
p_1 (T_{N,i}) + p_2 (T_{N,i}) = 1 - p_0 (T_{N,i}) \equiv q_1 + q_2
\end{equation}
where 
\begin{equation}
q_1 \equiv \frac {1}{N}\left[ n_0 g_1 + n_1 g_2 + n_2 g_0 \right]; \quad 
q_2 \equiv \frac {1}{N}\left[ n_0 g_2 + n_1 g_0 + n_2 g_1 \right] 
\end{equation}   
so that we may identify $p_1 (T_{N,i})$ either with $q_1$ or with $q_2$. That is to say, we may either have $p_1 (T_{N,i}) \equiv q_1 \equiv p_0 (T^1 _{N,i})$ or have 
$p_1 (T_{N,i}) \equiv q_2 \equiv p_0 (T^2 _{N,i})$, where we define the 
$\alpha$-complementary of $T_{N,i}$ as $T^\alpha _{N,i} \equiv t^\alpha _{i-N}\cdots 
t^\alpha _{i-1}$, with $t^\alpha \equiv (t + \alpha )($ mod $3)$. We choose the first definition:
\begin{eqnarray}
p_1 (T_{N,i}) & \equiv & p_0 (T^1 _{N,i}) 
                       =  \frac{1}{N} (n_0 g_1 + n_1 g_2 + n_2 g_0)\\
&&~~~\nonumber \\
p_2 (T_{N,i}) & \equiv & p_0 (T^2 _{N,i}) 
                       =  \frac{1}{N} (n_0 g_2 + n_1 g_0 + n_2 g_1)
\end{eqnarray}
Clearly, these definitions, Eqs.(1, 4 \& 5), ensure that 
$p_0 (T_{N,i})+ p_1 (T_{N,i})+  p_2 (T_{N,i}) = 1$. The other choice for $p_1 (T_{N,i})$
can be shown to lead to the same result with a simple coordinate transformation.
Taking $0$ and $1$ as the independent symbols, we can write Eqs.(1 \& 4)
in the form,
%\begin{widetext}
\begin{equation}
p_0 (T_{N,i})  \equiv  p_0 (n_0 ,n_1 ;N) 
                        =  \frac{1}{3} \left( 1-\frac{{\cal N}_0 }{N}\mu _{0} - \frac{{\cal N}_1}{N}\mu _{1}\right)
\end{equation}
\begin{equation}
p_1 (T_{N,i}) \equiv  p_1 (n_0 ,n_1 ;N)
              = \frac{1}{3} \left( 1+\frac{{\cal N}_1}{N}\mu _{0} + \frac{[{\cal N}_1 - {\cal N}_0)]}{N}\mu _{1}\right)
\end{equation}
%\end{widetext}
where ${\cal N}_0 \equiv [N - (2n_0 + n_1 )]$ and ${\cal N}_1 \equiv [N - (n_0 + 2n_1 )]$.

A ternary sequence of $N$ symbols is completely described by the probability, $Q(n_0 ,n_1 ;N)$, that there are $n_0$ number of zeroes and $n_1$ number of ones in the sequence:
\begin{eqnarray}
Q(n_0 ,n_1 ;N+1) & = & p_0 (n_0 - 1,n_1 ;N)Q(n_0 - 1,n_1 ;N) \nonumber \\
&& + p_1 (n_0 ,n_1 - 1;N)Q(n_0 ,n_1 - 1;N) \nonumber \\ 
&& + p_2 (n_0 ,n_1 ;N)Q(n_0 ,n_1 ;N)
\end{eqnarray}
Since the average number of any symbol in the sequence will be $N/3$ asymptotically ({\it i.e.,} no global bias in the system), we rewrite the above equation in terms of the variables, $x \equiv 3n_0 - N$ and $y \equiv 3n_1 - N$. In doing so, we make use of the correspondence 
$(n_0 ,n_1 ;N+1) \to (x-1,y-1;N+1)$, $(n_0 - 1,n_1 ;N) \to (x-3,y;N)$ 
and $(n_0 ,n_1 - 1;N) \to (x,y-3;N)$ as can be seen from the definitions of $x$ and $y$. 
\begin{eqnarray}
Q(x-1,y-1;N+1) & = & p_0 (x-3,y;N)Q(x-3,y;N)\nonumber \\
&& + p_1 (x,y-3;N)Q(x,y-3;N) \nonumber \\ 
&& + (1-[p_0 (x,y;N) + p_1 (x,y;N)])Q(x,y;N)
\end{eqnarray}
where the probabilities $p_0 (x,y:N)$ and $p_1 (x,y;N)$ are given by,
\begin{eqnarray}
p_0 (x,y;N) & = & \frac{1}{3} \left( 1+\frac{[2\mu _{0}+\mu _{1}]}{3N}x + 
                                        \frac{[\mu _{0}+2\mu _{1}]}{3N}y  \right) \quad \\
&& \nonumber \\
p_1 (x,y;N) & = & \frac{1}{3} \left( 1+\frac{[\mu _{1}-\mu _{0}]}{3N}x - 
                                        \frac{[2\mu _{0}+\mu _{1}]}{3N}y  \right)
\end{eqnarray}
The continuum version of Eq.(9) is then given by
%\begin{widetext}
\begin{eqnarray}
\frac{\partial Q}{\partial N} & = & \frac{1}{4}D\left( \frac{\partial ^2 Q}{\partial x^2 } + \frac{\partial ^{2}Q}{\partial y^2 }\right)  
- \frac{\gamma}{N + N_0 } \left( \frac{\partial [(\lambda _{0}x +\lambda _{1}y) Q]}{\partial x} +
\frac{\partial [(\lambda x -\lambda _{0}y) Q]}{\partial y}\right)
\end{eqnarray}
%\end{widetext}
where $D$ and $\gamma$ represent the diffusion and drift constants respectively. We have introduced the parameter $N_0$ to allow for the possible existence of transient time in the problem. The $\lambda$'s are defined by,
\begin{equation}
\lambda _0 = \frac{1}{3}(2\mu _0 + \mu _1 );\, 
\lambda _1  = \frac{1}{3}(\mu _0 + 2\mu _1 );\,
\lambda  = \frac{1}{3} (\mu _1 - \mu _0 ) 
\end{equation}

A standard method of solving Eq.(12), with the initial condition 
$Q(x,y;N=0)=\delta (x)\delta (y)$, is to first Fourier transform it with respect to the variables $x$ and $y$:
%\begin{widetext}
\begin{equation}
\frac{\partial {\tilde Q(q_x ,q_y ;N)}}{\partial N} = -\frac{1}{4} D(q_x ^2 + q_y ^2){\tilde Q} 
+ \frac{\gamma }{N + N_0 }\left( [\lambda _0 q_x + \lambda q_y ]
 \frac{\partial {\tilde Q}}{\partial q_x } 
+ [\lambda _1 q_x - \lambda _0 q_y ]\frac{\partial {\tilde Q}}{\partial q_y }\right)
\end{equation}
%\end{widetext}
where $q_{x}$ and $q_{y}$ are the Fourier conjugates of $x$ and $y$
respectively. This first order equation can then be solved by the method
of characteristics. In particular, we have to solve the equations,
\begin{equation}
\gamma\frac{dN}{N + N_0 } = \frac{dq_x }{\lambda _0 q_x + \lambda q_y }                                         
                         = \frac{dq_{y}}{\lambda _{1}q_{x}-\lambda _{0}q_{y}}
\end{equation}
Considering the second and the third terms leads to the equation,
\begin{equation}
(-\lambda _0 q_y + \lambda _1 q_x )dq_x = (\lambda _0 q_x + \lambda q_y )dq_y
\end{equation}
that can be solved for $q_x$ in terms of $q_y$ or {\it vice versa}:
\begin{equation}
q_x = \frac{(\alpha + 2\lambda _0 )}{\lambda _1 }q_y ; \quad 
\mbox{or}\quad q_y = \frac{\alpha}{\lambda}q_x; \quad \mbox{where}\quad 
\alpha \equiv -\lambda _0 \pm [\lambda _0 ^2 + \lambda \lambda _1 ]^{1/2}
\end{equation}
Using these relations, we can immediately obtain their $N$-dependence.
\begin{equation}
q_x ,\, q_y \propto (N + N_0 )^{ \gamma \Gamma }; \quad \Gamma \equiv \pm [ \lambda _0 ^2 + \lambda \lambda _1 ]^{1/2} 
\end{equation}
which in turn helps us define the variables,
\begin{equation}
\xi \equiv q_x (N + N_0 )^{\Gamma};\quad \eta \equiv q_y (N + N_0 )^{\Gamma}
\end{equation}
in terms of which Eq.(14) can be written as
\begin{equation}
\frac{\partial { \tilde Q(\xi ,\eta ;\tau )}}{\partial \tau} = -\frac{1}{4}D(\xi ^2 + \eta ^2 )\tau ^{-2\Gamma} { \tilde Q(\xi ,\eta ;\tau )}
\end{equation}
where $\tau \equiv (N + N_0 )$. With the initial condition, 
${ \tilde Q(q_x ,q_y ;\tau = N_0 )}=1$, we immediately have the solution,
\begin{equation}
\tilde Q(q_x ,q_y ; \tau ) 
= \exp \left \{- \frac{1}{4}\sigma ^2(\tau)(q_x ^2 + q_y ^2 )\right \}
\end{equation}
from which we identify the variance,
\begin{equation}
\sigma ^2(\tau) = \left[ 1-\left( \frac{N_0 }{N + N_0 }\right)^{-2\Gamma +1}\right] \frac{\tau}{(1-2\Gamma)}; \quad 2\Gamma \neq 1
\end{equation}
It is clear from the above expression that $\sigma ^2 (\tau) \propto \tau ^{\nu}$,
where $\nu =1$ whenever $2\Gamma < 1$ and $\nu =2\Gamma$ whenever
$2\Gamma > 1$. Even in the case, $2\Gamma = 1$, the exponent $\nu =1$
but there will be logarithmic corrections.

The existence of a critical value, $\Gamma _c = 1/2$, for $\Gamma$ implies that the parameter space,
$ \{(\mu _0 ,\mu _1 )\mid -1\leq \mu _0 ,\mu _1 \leq 2 \, \& \, -2 \leq (\mu _0 + \mu _1 )\leq 1 \}$,
is divided into two regions, diffusive and superdiffusive. As shown in Fig.1, the diffusive region defined by the condition $\Gamma \leq 1/2$ is the elliptical region, 
$\mu _{0}^{2}+\mu _{1}^{2}+\mu _{0}\mu _{1} \leq 3/4$, inscribed within the triangular phase-space. Interestingly, the area of this diffusive region is $\pi \sqrt{3}/2$ which is roughly 
$60 \% $ of the available phase-space area. This may be contrasted with the binary case where it is exactly $50 \% $. Now the question arises whether a diffusive subregion of the ternary is likely to be mapped into a superdiffusive region of the binary due to a coarse-graining process.
Since a mapping of a set of three symbols to a set of two symbols always introduces a global bias in the resulting binary system, it is necessary to reformulate the binary case so that we can identify long-range correlation in a biassed sequence. 
%
% FIGURE, phase2d.eps
%
\begin{figure}
\includegraphics[width=3.25in,height=2.75in]{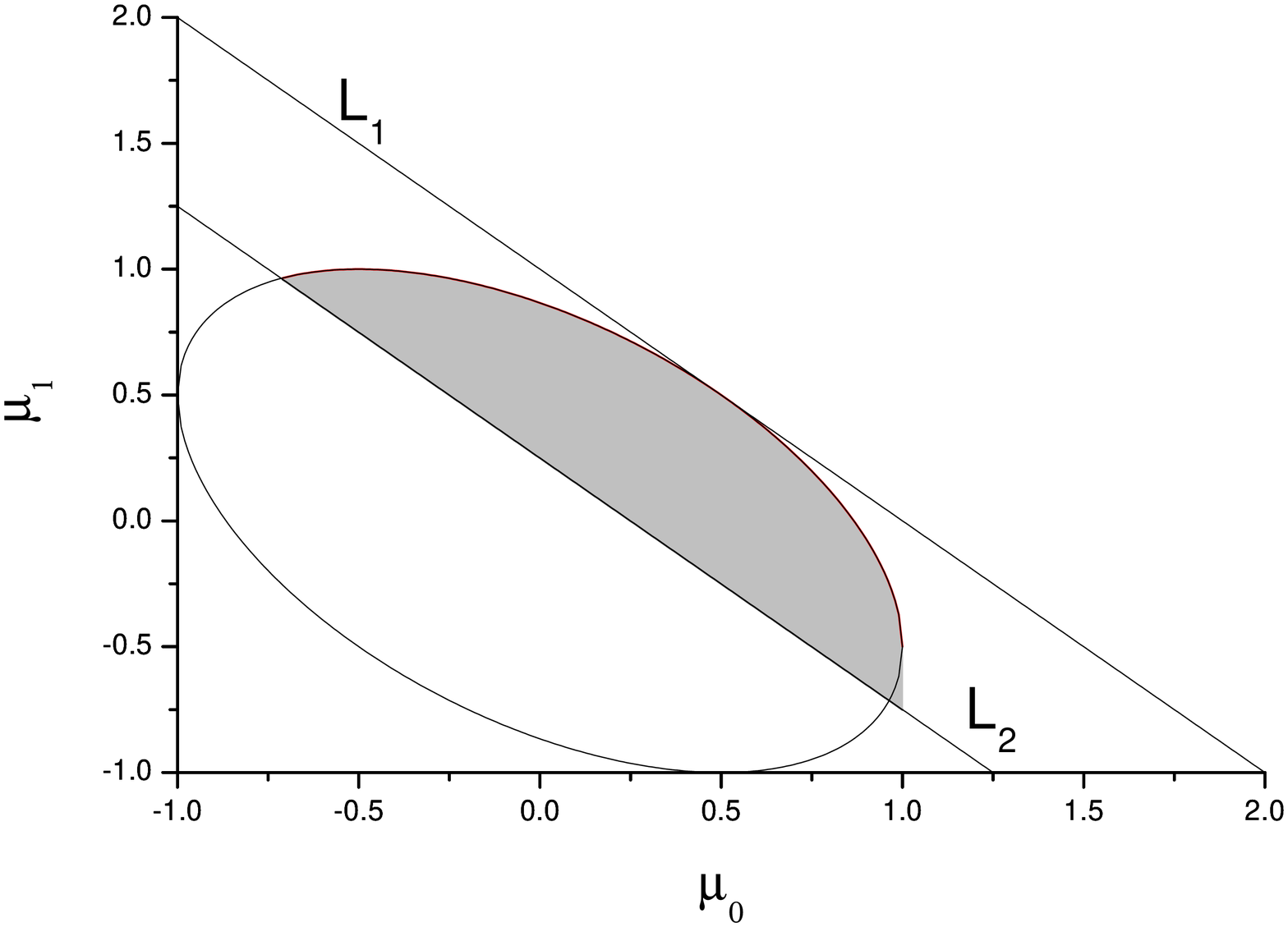}
\caption{Phase diagram for the ternary sequence. $L_1$ and $L_2$ refer to the lines $\mu _0 + \mu _1 = 1$ and $\mu _0 + \mu _1 = 1/4$ respectively. The ellptical region of the triangular phase space corresponds to the {\it diffusive} behavior whereas the region exterior to it corresponds to the 
{\it superdiffusive} behavior of the ternary sequences. The entire region between the lines $L_1$ and $L_2$, inclusive of the shaded elliptical region, gets mapped into the superdiffusive regime of the binary.}
\end{figure}

\section{Binary Sequences, revisited} 
The conditional probability, $p_0 (n_0 ,N)$, of finding zero as the $(N+1)^{th}$ symbol in a binary sequence that already consists of $n_0$ zeros is given by the definition,
\begin{equation}
p_0 (n_0 ,N)  \equiv  \frac{1}{N} \left( n_0 g_0 + [N - n_0 ]g_1 \right) 
\end{equation}
where $g_0$ and $g_1 = 1 - g_0$ are the {\it a priori} probabilities of choosing symbols zero and one respectively. If a biassed coin ({\it i.e.,} $g_0 \neq 1/2$) is used for building up a sequence, then the average number of zeros in the sequence, $<n_0>$, is expected to be 
$b N$ asymptotically, where the intrinsic bias $b$ is equal to $g_0$ if and only if the symbols are added blindly without reference to the existing sequence ({\it i.e}., $p_0(n_0, N)$ does not depend on $n_0$). In order that the equality $<n_0> = b N$ holds good in general, it is necessary that the dependence of $p_0(n_0, N)$ on $n_0$ is through the variable, $x \equiv (n_0/b) - N$. In analogy with the unbiassed case [4], we write $g_0 = b(1 + \mu)$ where $\mu \in [-1, -1 + (1/b)]$ pametrizes the deviation from the intrinsic bias of the coin. We then have
\begin{equation}
p_0 (x,N) = \alpha + \frac{\beta}{N+N_0}~x \quad \mbox{where}\quad \beta \equiv b(2g_0 - 1);
\quad \alpha \equiv \beta - g_0 + 1 
\end{equation}
In order to check whether the equality $<n_0> = b N$ holds good for $\mu$ in the range $[-1, -1 + (1/b)]$, we have generated [8] a large number of sequences each consisting of upto hundred thousand symbols for $b=2/3$. We find that $<n_0>/N \approx 0.655$ for $\mu = 1/2$ and decreases fast to the value $1/2$ for $\mu \leq 0.3$. On the other hand, the variance $\sigma ^2(N) \equiv <n_0^2> - <n_0>^2$ turns out to be proportional to $N^{2(2g_0 - 1)}$ for $g_0 > 3/4$, and to $N$ for $g_0 \leq 3/4$. This implies that $\alpha$, defined as above, does not lead to the expected constant value for $<n_0>/N$, whereas $\beta$ correctly defines the correlation exponent. Since the above definition of $\alpha$ is derived from Eq.(23) for $p_0(n_0, N)$, we have to check whether Eq.(23) is a valid definition also for the biassed case. Treating $\alpha$ as a parameter to be fixed later, we can easily show that the probability, $Q(n_0 ,N)$, that there are $n_0$ zeros in the sequence satisfies the following equation in the continuum limit:
\begin{equation}
\frac{\partial Q}{\partial N} = \frac{1}{2b^2}(1 - b^2)\frac{\partial ^2 Q}{\partial x^2}
+ (1 - \frac{\alpha}{b})\frac{\partial Q}{\partial x} 
- \frac{\beta}{b(N + N_0)}\frac{\partial [xQ]}{\partial x}
\end{equation}
Solving this equation, we can show that the distribution, $Q(n_0,N)$, peaks at $<n_0> = (1 + f(\alpha))bN$ asymptotically where $f(\alpha) \equiv (b - \alpha)/[2b(g_0 - 1)]$. Thus, the peak will remain fixed at $bN$ for all values of $\mu$ only if $\alpha = b$. This leads to the following definition for $p_0(n_0, N)$:
\begin{equation}
p_0(n_0, N) \equiv b\left(1 + \frac{(2g_0 - 1)}{N}\left[\frac{n_0}{b} - N\right] \right)
= 2b(1 - g_0) + (2g_0 - 1)\frac{n_0}{N}
\end{equation}
It is a general definition applicable even to an unbiassed sequence ({\it i.e.,} reduces to the definition, Eq.(23), for $b = 1/2$). The variance of the distribution is given by,
\begin{equation}
\sigma ^2 (N) \propto \left\{\begin{array}{lll}
                              N + N_0 & \mbox{for} & (2g_0 - 1)\leq 1/2 \\
                             (N + N_0)^{2(2g_0 - 1)} & \mbox{for} & (2g_0 - 1) > 1/2
                             \end{array}
                      \right.
\end{equation}
Since $g_0 = b(1 + \mu)$, long-range correlation in the sequence will be characterized by the values of $\mu$ in the range $\mu \in (-1 + [3/4b], -1 + [1/b])$. Again, we have numerically checked [8] that the condition $\alpha = b$ ($ = 2/3$ in our case) does ensure the constancy of $<n_0>/N (=2/3)$ and, more importantly, we have confirmed that the variance has the above expected behaviour. In other words, the LRC behavior of a binary sequence is characterized by the exponent, $(2g_0 - 1)$, for $g_0 > 3/4$.  

\section{Mapping the ternary into the binary}
Assume that the symbols zero and one of the ternary are identified as '$0$' while the symbol two is identified as '$1$'. Then the {\it a priori} probability for '$0$' and '$1$' will be
\begin{equation}
g_0 = \frac{1}{3} (2 + \mu _0 + \mu _1 );\quad g_1 = \frac{1}{3} (1 + \mu _2 )   
\end{equation}
The condition for non-trivial correlations then turns out to be $\mu _0 + \mu _1 > 1/4$, above
the lower line shown in Fig.1. The entire region of the phase space between this line and the
upper line, $\mu _0 + \mu _1 = 1$, corresponds to long-range correlation when mapped into the biassed binary. This is so because the correlation parameter, $\mu$, for the binary
is related to $\mu _0$ and $\mu _1$ by the identity, $\mu = (\mu _0 + \mu _1)/2$
which in turn implies a correspondence between the range $1/4 \leq (\mu _0 + \mu _1) \leq 1$ of the ternary and the range $1/8 \leq \mu \leq 1/2$ of the biassed binary; every line in between and parallel to the upper and lower lines in Fig.1 is mapped into a point in the range $1/8 \leq \mu \leq 1/2$ and {\it vice versa}. In particular, the shaded part of the ellipse in Fig.1 is the diffusive area that is mapped into the superdiffusive regime of the biassed binary, which in general is characterized by a parameter, $\mu$, in the range $\mu \in (-1 + [3/4b], -1 + [1/b])$.

Conversely, the fact that the correlation parameter of the biassed binary sequence under study has a value in the range $\mu \in (-1 + [3/4b], -1 + [1/b])$ does not necessarily mean that the parent ternary sequence also has long-range correlations. Even if we know {\it a priori} that there are long-range correlations between symbols of the parent sequence, the parameter $\mu$ is not a true measure of its strength.

\section{Summary}
We have worked out the exact phase diagram for a ternary sequence with long-range correlation. Motivated by the fact that the diffusive portion of the phase space is larger for the ternary than for the binary, we have studied the mapping between these two. We have shown that long-range correlation for the binary does not necessarily imply long-range correlation for the ternary. This exact result has deeper implications for the coarse-graining of many-alphabets sequences [8]. For example, if we do not know that the original sequence has long-range correlations or (and) if we do not know the coarse-graining procedure that has led to the representative sequence under study, then we may not be able to make accurate inferences about the correlation properties of the original sequence. A systematic numerical study of this problem will be reported elsewhere. It could be of interest to do a similar study for a non-Markov sequence.  

\begin{acknowledgements}
We wish to thank the Referee for fruitful comments. SLN acknowledges discussions with Y. S. Mayya, S. V. Suryanarayan and P. S. R. Krishna. 
\end{acknowledgements}

\vspace{1pc}
$^\dagger$slnoo@magnum.barc.ernet.in

\end{document}